\documentclass[aps,preprint,nofootinbib,letterpaper]{revtex4}%
\usepackage{amssymb}
\usepackage{amsfonts}
\usepackage{amsmath}
\usepackage{amsmath}
\usepackage{graphicx}
\usepackage{silence}
\usepackage[hyperfootnotes=false]{hyperref}

\usepackage[usenames]{color}%
\providecommand{\U}[1]{\protect\rule{.1in}{.1in}} 
\hypersetup{colorlinks,linkcolor={blue},citecolor={blue},urlcolor={black}} 
\providecommand{\U}[1]{\protect\rule{.1in}{.1in}}
\definecolor{blue}{rgb}{0,0,1}

\definecolor{red}{rgb}{1,0,0}

\begin{document}
\title{A smooth road to bumpy horizons: shaping black holes with non-linear sigma models, from supergravity to higher dimensions}
\author{Fabrizio Canfora$^1$, Nicolás Grandi$^2$, Carla Henríquez-Báez$^3$, Julio Oliva$^4$}
\affiliation{$^{1}$Centro de Estudios Científicos (CECS), Casilla 1469, Valdivia, Chile and Facultad de Ingeniería, Universidad San Sebastián, General Lagos 1163, Valdivia 5110693, Chile.}
\affiliation{$^{2}$Instituto de Física La Plata, CONICET and Departamento de Física Dr. Emil Bose, UNLP Casilla de correos 67, 1900 La Plata, Argentina. }
\affiliation{$^{3}$Centro Multidisciplinario de Física, Vicerrectoría de Investigación, Universidad Mayor, Camino La Pirámide 5750, Santiago, Chile.}
\affiliation{$^{4}$Departamento de F\'{\i}sica, Universidad de Concepci\'{o}n, Casilla 160-C, Concepci\'{o}n, Chile.}

\begin{abstract}
We construct new families of solutions for General Relativity coupled to a general class of non-linear sigma models, some of which can be embedded in supergravity. The solutions include neutral, charged and magnetized black holes with bumpy horizons, bumpy stars, and anisotropic cosmologies in $d\geq 4$ dimensions, as well as black strings and black $p$-branes. We also present a family of time-dependent solutions in $2+1$-dimensions. The construction relies on a set of first-order Bogomol'nyi-Prasad-Sommerfield relations for the coset scalars, that were recently exploited for the construction of bumpy black holes on the non-linear sigma model with homogeneous target $SU(2)/U(1)$ in 2601.22914 [hep-th].
\end{abstract}

\maketitle

\section{Introduction}
Non-linear sigma models (NLSMs) play a fundamental role in physics. They emerge in the effective description of pions through chiral perturbation
theory (or $\chi$PT, see e. g. \cite{CHPT1, CHPT2, CHPT3, CHPT4} and references therein). Moreover, they also appear both in dimensional reductions of General Relativity, allowing to identify hidden symmetries of highly symmetric sectors of the theory \cite{Belinski:2001ph}, as well as in String Theory \cite{Green:2012oqa}. This class of models play a fundamental role in ungauged and gauged  supergravities, describing the dynamics of the scalar sector (which may span a general, a coset, a homogeneous or a target space with more geometric structure, see e.g. \cite{Trigiante:2016mnt} and references therein). In the present manuscript, we want to analyze how the NLSM dynamics determines a fundamental feature of black holes, black strings and black branes, namely the geometry of the horizon. We concentrate on a suitable Bogomol'nyi-Prasad-Sommerfied (BPS) sector of the model. 

Assuming the dominant energy condition is satisfied, the horizon topology of asymptotically flat, four-dimensional, stationary black holes is fixed to that of the two-sphere \cite{Hawking:1971vc}. The inclusion of a negative cosmological constant allows to construct black holes with compact two-dimensional horizon surfaces of arbitrary genus, at the cost of propagating such non-trivial topology to the asymptotic region, giving rise to topological black holes which in general, are asymptotically locally AdS \cite{Lemos:1994xp,Mann:1996gj,Vanzo:1997gw,Brill:1997mf}. In all of these cases, the horizon has a constant local curvature,  obtained by a quotient of the maximally symmetric space of the given sign of curvature, which in general reduces the number of Killing vectors, even leading to horizons with constant local curvature without Killing vectors.

Recently  in the work \cite{Canfora:2026col}, it has been shown that the NLSM on a target space given by a two-sphere (or on the equator of the three-sphere if one is willing to emphasize the $SU(2)$ origin of the construction), lead to new static black holes within the simple family 
\begin{equation}\label{funosobref}
    ds^2=-f(r)dt^2+\frac{dr^2}{f(r)}+r^2d\Sigma_2^2\ ,
\end{equation}
with the local geometry of the horizon manifold $\Sigma_2$ determined by the scalar fields. There, the black holes were interpreted as having {\em bumpy} horizons supported by superfluid pions. The latter statement is motivated by the fact that the matter field equations are solved by a BPS-like first order system, which implies the second order one, and allows for the superposition of the bumps. Parametrizing the horizon in stereographic coordinates, the logarithm of the conformal factor fulfills a non-homogeneous Liouville equation that is sourced by the matter fields. 
The non-homogeneous term is determined by the specific NLSM considered in \cite{Canfora:2026col}, and the matter sector is solved by flat-harmonic functions of the horizon coordinates.

In this work, we extend the results of \cite{Hawking:1971vc} for general NLSMs. This allows to explore the embedding of the present spacetimes in gauged and ungauged supergravity, leading in the simplest cases to a conformal factor which is completely determined by the Kähler potential of the two-dimensional target space metric. 

In Section \ref{sec:NLSM}, we consider a two-dimensional NLSM in a conformally flat parametrization on the target space. Remarkably, the Einstein equations turn out to imply a simple non-linear equation for the conformal factor of the horizon metric. For the NLSM scalar, the Einstein equations enforce a set of Bogomol'nyi-Prasad-Sommerfield (BPS) relations, which in turn imply the scalar equations. This permits to generalize the bumpy black holes of \cite{Canfora:2026col}, to more general NLSMs. Section \ref{sec:Kahler}, the results are naturally extended by including a complex parametrization of both, the target space and the horizon geometry. For a particular choice of parameters, the logarithm of the conformal factor equals the Kähler potential of the target space metric. The complex scalar field turns out to be a holomorphic function of the coordinates of the horizon. In Section \ref{sec:additional.sources} we explore three different extensions of these bumpy solutions. First we investigate the compatibility of the equations under the inclusion of more scalar fields. This will permit the construction of boson star and other compact objects, which usually require a numerical integration. Then, we construct the four-dimensional, dyonic solution by introducing a minimally coupled Maxwell field. The equations for the bumpy black holes supported by the NLSM turn out to be perfectly compatible with both an electric and a magnetic ansatz. We finish Section \ref{sec:additional.sources} exploring the consistency of the BPS sector of the gravitating NLSM equations, with a generic matter field described by a the energy-momentum tensor of a perfect fluid. This will permit exploring the formation of the bumpy black holes via gravitational collapse. Section \ref{sec:higher.dimensions} extends the result to higher dimensions via different avenues. The inclusion of $n$ NLSM doublets, permits constructing bumpy black holes in higher dimensions with horizons given by the direct product of $n$ two-dimensional Euclidean spaces. The lapse function is given by the one of the Schwarzschild-Tangherlini spacetime, but the geometry can be more general than a constant curvature space or a $d-2$-dimensional Einstein manifold. We also show that, even with a fixed number of NLSM doublets, the horizon geometry can be extended by adding an Einstein manifold properly multiplied by the square of the area function. Such Einstein manifold can be flattened if one extends the dimension of the target space with flat directions. These new scalars turn out to be proportional to the coordinates parametrizing the new flat directions of the horizon. We finish Section \ref{sec:higher.dimensions} with the construction of black strings and black $p$-branes, which is non-trivial since the in the presence of matter fields in four-dimensions it is well-known that there are obstructions to trivially extend the four-dimensional solutions, both when the cosmological constant vanishes and when it does not. We show that the consistency of the equations permits to construct black strings and $p$-branes via simple compactifications, or via the inclusion of a suitable brane warp-factor depending on the coordinates of the extended directions, in the presence of a cosmological constant. Section \ref{sec:time.dependent} is devoted to the study of time-dependent solutions first in the context of non-homogeneous cosmologies in four and three dimensions, as well as the proof of a Birkhoff's like theorem. Section \ref{sec:conclusions} contains conclusions and further comments. 
\section{Einstein gravity coupled to a non-linear sigma model}
\label{sec:NLSM}
Let us first consider Einstein gravity with a cosmological constant coupled to a NLSM, which play an important role  
in physics. Considering first a simple two-dimensional target space spanned by the scalars $\alpha(x^\mu)$ and $\phi(x^\mu)$, the action can be written as
\begin{equation}\label{EHNLSM}
I_{\sf EH-NLSM}=\int d^{4}x\sqrt{-g}\left(\frac{R-2\Lambda}{2\kappa}  -\frac{1}{2}\left(  \nabla
\alpha\right)  ^{2}-\frac{1}{2}\Omega^2(  \alpha)   \left(
\nabla\phi\right)  ^{2} -V\left(  \alpha,\phi\right) \right)  \ .
\end{equation}
In the reference \cite{Canfora:2026col} the scalar correspond to the Euler angles parametrization of the $SU(2)$ NLSM, restricted to the equator of the three-sphere group manifold, which results in a function $\Omega(\alpha)=\sin\alpha$. 
In such pionic case, $\Omega(\alpha)$ 
 represents the amplitude of the pionic superfluid wave function while the gradient of $\phi$ represents the superfluid velocity (see the discussion in  \cite{CHPT5} and \cite{CHPT6}).
In the present section, we will keep $\Omega(\alpha)$ generic, covering a family of NLSMs characterized by this arbitrary function generalizing substantially the results of \cite{Canfora:2026col}. Such scalar Lagrangian accommodates many relevant models that span from supergravity to pion physics, as mentioned in the introduction.

The above action allows for a very simple field redefinition, given by $d\chi=d\alpha/\Omega(\alpha)$, which turns it into the somewhat simpler expression
\begin{equation}\label{eq:action.chi}
I_{\sf EH-NLSM}=\int d^{4}x\sqrt{-g}\left(\frac{R-2\Lambda}{2\kappa}  -\frac{1}{2}\Omega^2(\chi)\left(  (\nabla \chi)^{2}+(
\nabla\phi)  ^{2}\right)  -V\left(  \chi,\phi\right)  \right)  \ ,
\end{equation}
where we see that $\Omega(\chi)=\Omega(\alpha(\chi))$ is the conformal factor of the metric on the two-dimensional sector of target space spanned by the fields $\phi$ and $\chi$.
For the particular case studied in  \cite{Canfora:2026col}, we have $\chi=\log\tan\frac\alpha2$ and $\Omega(\chi)={\rm sech}\,\chi$.

With the form \eqref{eq:action.chi} of the action, the matter field equations take the simple form
\begin{align}\label{laeqparaalpha}
&\nabla_{\mu}\!\left(  \Omega^2(\chi)\partial^{\mu}\chi\right)
-\frac{\partial V   }{\partial\chi}=\Omega(\chi)\frac{\partial \Omega}{\partial\chi}( (\partial \chi)^{2}+(
\partial\phi)  ^{2})\ , 
\quad
&\nabla_{\mu}\!\left(  \Omega^2(\chi)\partial^{\mu}\phi\right)
-\frac{\partial V   }{\partial\phi}=0\ . 
\end{align}
On the other hand, Einstein equations read
\begin{equation}\label{einsteinequations}
G_{\mu\nu}+\Lambda g_{\mu\nu}=\kappa \,T^{\text{\sf NLSM}}_{\mu\nu}\ ,
\end{equation}
where the stress-energy tensor takes the form
\begin{equation}
T^{\text{\sf NLSM}}_{\mu\nu}=\Omega^2(\chi)\left( \partial_{\mu}\chi\partial_{\nu}\chi+\partial_{\mu
}\phi\partial_{\nu}\phi-\frac{1}{2}g_{\mu\nu}\left((
\nabla\chi)  ^{2}+\left(  \nabla\phi\right)
^{2}\right)\right)  -g_{\mu\nu}V\left(  \chi,\phi\right)\ .
\end{equation}

We want to consider the above defined model on a manifold of the form
\begin{equation}\label{lametrica}
ds^{2}=-f(r)N^2(r)  dt^{2}+\frac{dr^{2}}{f(r) }%
+r^{2}d{\Sigma}_2^{2} \ .
\end{equation}
Here the manifold $\Sigma_2$ is Euclidean and two-dimensional, but otherwise arbitrary. 
In the case in which the manifold $\Sigma_2$ has constant curvature, the above ansatz reduces to the topological extension of the Schwarzschild solution with constant curvature horizons \cite{Lemos:1994xp,Mann:1996gj,Vanzo:1997gw,Brill:1997mf}. However, a fundamental question on General Relativity, which may have astrophysical consequences, can be stated: 
is it always the case that black holes with realistic matter sources possess horizons with constant curvature? Or in other words:  under which conditions can black hole horizons be ``bumped'' {\em i.e.} have a non-constant curvature?  
Moreover, when this happens, which is the physical mechanism which prevents the bumps of the horizon from being stretched by the  
gravitational field?  The main goal of the present manuscript is to contribute to the answer to such questions. Of course when rotation is included one may expect departures from the horizons of local constant curvature, as it happens in the vacuum case, that's why here we are addressing the previous questions in a static family of spacetimes \eqref{lametrica}.

We parametrize the transverse manifold $\Sigma_2$ with coordinates $x,y,$ in such a way that its metric is manifestly conformally flat $d\Sigma_2=e^{P(x,y)}\left(dx^2+dy^2\right)$ where $P(x,y)$ is an arbitrary function. We can then write 
\begin{equation}
ds^{2}=-f(r) N^2(r) dt^{2}+\frac{dr^{2}}{f(r)  }%
+r^2e^{P(x,y)}\left(dx^2+dy^2\right)\ .
\label{eq:metric.Ansatz}
\end{equation}

It is well-known that in vacuum Einstein theory with a cosmological constant, the Ansatz \eqref{eq:metric.Ansatz} leads to topological black holes, where the manifold $\Sigma_2$ must have a constant curvature ${R}_{\Sigma_2}=2\gamma$  and the metric functions reduce to 
\begin{equation}
f(r)=\gamma-\frac{2m}{r}-\frac{\Lambda}{3}r^2\ ,
\qquad\qquad\qquad N(r)=1\ .
\label{eq:vaccum.solution}
\end{equation} 
The family of solutions then contain asymptotically flat or dS spherical black holes, and spherical, planar and hyperbolic black holes which are asymptotically locally AdS, depending on the value of the cosmological constant. Here, inspired by the recent results of \cite{Canfora:2026col}, we want to investigate which types of deformation of $\Sigma_2$ may be induced by the scalars $\phi,\chi$ depending only on the directions $x,y$, when the function $\Omega(\chi)$ is kept arbitrary. This broadens the family of NLSMs beyond the one considered in \cite{Canfora:2026col}. 

Let us first address Einstein equations:
\small
\begin{align}
&(tt)&&
\!\!-\frac{fN^2}{r^2}\left(f+rf'+r^2\Lambda-\frac12 R_{\Sigma_2}\right)= \frac{\kappa\,fN^2}{2r^2}\left(e^{-P}\Omega^2(\chi)\left((\partial\chi)^2+(\partial\phi)^2\right)+2r^2V \right) \ ,
\nonumber\\ 
&(rr)&&
\frac{1}{r^2f}\left(f+rf'+r^2\Lambda+2rf\frac {N'}N-\frac12 R_{\Sigma_2}\right)= -\frac{\kappa}{2r^2f}\left(e^{-P}\Omega^2(\chi)\left((\partial\chi)^2+(\partial\phi)^2\right)+2r^2V \right)\ ,
\nonumber\\ 
&(xx)&&\kappa\,
re^P\left(rf\frac{N''}{N}+\frac{N'}N\left(f+\frac32rf'\right)+f'+\frac {rf''}2+r\Lambda\right)=
\nonumber
\\&&&\qquad \qquad \qquad \qquad \qquad \qquad \qquad
=\frac{\kappa\,}{2}\Omega^2\left((\partial_x\phi)^2-(\partial_y\phi)^2
+(\partial_x\chi)^2-(\partial_y\chi)^2\right)-e^Pr^2V \ ,
\nonumber\\ 
&(xy)&&
0=\kappa\,\Omega^2\left(\partial_x\chi\partial_y\chi+\partial_x\phi\partial_y\phi\right)\ .
\label{eq:Einstein.vaccuum}
\end{align}
\normalsize
In these equations, the notation $(\partial\, \cdot\,)^2$ denotes that the derivatives are contracted with the Euclidean flat two-dimensional metric, namely $(\partial\,  \cdot\,)^2=(\partial_x\, \cdot\,)^2+(\partial_y\, \cdot\,)^2$. Also  $R_{\Sigma_2}=-e^{-P}\partial^2P$ is the scalar curvature of the would-be horizon manifold $\Sigma_2$. 

The first observation is that the $(xy)$ equation imposes a constraint in the scalar fields. This is solved by requiring that the pair $\phi,\chi$ satisfies Cauchy-Riemann relations, 
\begin{equation}
    \partial_x\phi=\partial_y\chi\ ,\qquad\qquad\partial_y\phi=-\partial_x\chi\ .
    \label{eq:Cauchy-Riemann}
\end{equation}
Interestingly, these conditions make the derivatives of the scalar fields to cancel in the $(xx)$ equation. The resulting expression depends on $x,y,$ only through the scalar fields inside the potential. This equation is not separable, unless we restrict the class of models under consideration to those with vanishing potential\footnote{The compatibility of the $(xx)$ equation can be achieved in an ansatz that is the direct product of two-dimensional spaces.}. Doing that, we can immediately solve the $(xx)$ equation in \eqref{eq:Einstein.vaccuum} for $f$, obtaining exactly the vacuum expression \eqref{eq:vaccum.solution} where, up to this point, $\gamma$ emerges as a separability constant.

However, we still have to solve the $(tt)$ and $(rr)$ equations. Inserting the explicit form of $f$ into them, we get an equation for the function $P(x,y)$, in the form
\begin{equation}
     {R}_{\Sigma_2}  =2\gamma+\kappa\,e^{-P}\Omega^{2}(\chi)\left((\partial \chi)^2+(\partial \phi)^2\right) \ .
     \label{eq:horizon.curvature}
\end{equation}
For vanishing matter fields, we obtain the known vacuum result, since the horizon curvature is fixed to $2\gamma$. For non-vanishing matter fields instead, it is the scalar field profile what will determine the horizon curvature. Since in two dimensions, up to diffeomorphism, a metric is determined by a single function (for example, the conformal factor $P(x,y)$ in the conformally flat gauge we have chosen), equation \eqref{eq:horizon.curvature} suffices to determine locally the metric, once the matter fields are given. This can be written more explicitly as
\begin{equation}
     \partial^2P  +2\gamma e^{P } =-2\kappa\,\Omega^2(\chi)(\partial \chi)^2 \ .
     \label{eq:liouville}
\end{equation}
The latter is the non-homogeneous Liouville equation, where the source is given by squared gradient of the scalar field profile $\chi(x,y)$. Notice that this is exactly the same equation as in \cite{Canfora:2026col}, but now for a more general NLSM, beyond the $SU(2)/U(1)$ example. 
 
Let us now address the analysis of the matter equations. We have already proved that the potential $V(\chi,\phi)$ has to vanish, and imposed an ansatz for the metric and the matter fields. 
Under these conditions, the scalar equations reduce to
\begin{align}
\label{laeqparaalpha3}
&  \Omega(\chi)\partial^2\chi
 =\frac{\partial \Omega}{\partial\chi}\left( (
\partial\phi)  ^{2} -(
\partial\chi)  ^{2}\right)\ , 
\qquad
&  \Omega(\chi)\partial^2\phi
 =-2 \frac{\partial \Omega}{\partial\chi} (
\partial\phi\cdot \partial\chi)\ ,
\end{align}
where again derivatives are contracted with the flat Euclidean two-dimensional metric. 
These equations are identically satisfied when the pair $\phi,\chi$ satisfies Cauchy-Riemann relations \eqref{eq:Cauchy-Riemann}. 
As these imply that $\phi, \chi$ are both harmonic functions in flat $x,y$ space, the left hand sides of \eqref{laeqparaalpha3} vanish. On the other hand, direct substitution shows that the right hand sides of \eqref{laeqparaalpha3} are also zero. This implies that the Cauchy-Riemann relations \eqref{eq:Cauchy-Riemann} act as effective BPS conditions, extending the proposal of \cite{Canfora:2026col} to more general NLSMs. 

According to equation \eqref{eq:liouville}, the horizon curvature $R_{\Sigma_2}$ is completely determined by the harmonic function $\chi$. This results  in a ``bumpy'' horizon, namely a horizon with non-constant curvature. Notice that such bumpiness will propagate up to infinity. 

{It is worth emphasizing the following important point. The Cauchy-Riemann conditions in the present framework can be considered as a set of BPS equations for the matter fields. The relevance of such BPS conditions is that, in many situations, they ensure that the corresponding configurations cannot be deformed to the trivial ones due to topological conservation laws. In order for a bump to resist against the tendency of strong gravity to ``undo it'', the bump needs to have some mechanism which can protect it. The topological conservation laws typical of BPS configurations provide us with such a mechanism.}
\section{Generalizing the scalar field dynamics: Supergravity}
\label{sec:Kahler}
An interesting point to notice is that our action \eqref{eq:action.chi} can be written in terms of a complex field $\Psi=\chi+i\phi$ and a Kähler metric,  in the form 
\begin{equation}
\label{eq:action.SuGra}
I_{\sf EH-NLSM}=\int d^{4}x\sqrt{-g}\left(\frac{R-2\Lambda}{2\kappa}  -\frac{1}{2} K_{\Psi\bar\Psi}(\Psi, \bar\Psi) \,\nabla^\mu\Psi\nabla_{\mu}\bar\Psi   \right)  \ ,
\end{equation}
where we have defined the Kähler metric as
\begin{equation}
    K_{\Psi\bar\Psi}(\Psi, \bar\Psi) =\Omega^2\!\left(\frac{\Psi+\bar\Psi}2\right)
    \label{eq:shift.symmetry}
\end{equation}
which can be understood as the derivative of a Kähler potential  with respect to the the complex field $\Psi$ and its conjugate $\bar \Psi$, or $K_{\Psi\bar\Psi}(\Psi,\bar\Psi)=\partial_\Psi\partial_{\bar\Psi}K(\Psi,\bar\Psi)$. This embeds our model into any Supergravity theory with a minimally coupled scalar sector. 

The specific form \eqref{eq:shift.symmetry} of the Kähler metric, depending on the sum $\Psi+\bar\Psi$, contains the class of metrics satisfying the ``no-scale'' condition, which appears in the context of string compactifications \cite{Ibanez:2012zz}. It manifests an imaginary shift symmetry $\Psi\to \Psi+i \Delta$, where $\Delta$ is any real spacetime function. 

The scalar equations in this new language read  
\begin{equation}
K_{\Psi\bar\Psi} \,{\square\Psi}  +K_{\Psi\Psi\bar\Psi} (\nabla\Psi)^2  = 0\ ,
\end{equation}
which is understood as the imposing equation and its complex conjugate. Defining the complex coordinate $\zeta=x+iy$ and limiting the dependence on the scalar to $\Psi(\zeta,\bar\zeta)$ we get
\begin{equation}
K_{\Psi\bar\Psi} \,{\partial_\zeta\partial_{\bar\zeta}\Psi}  +K_{\Psi\Psi\bar\Psi} \partial_\zeta\Psi\partial_{\bar\zeta}\Psi  = 0\ .
\label{eq:scalar.complex}
\end{equation}
In this setup, the Cauchy-Riemann condition, which ensures the closure of Einstein and scalar equations, translates into the holomorphicity of the complex field $\Psi$ as a function of the complex variable $\zeta$, namely $\Psi=\Psi(\zeta)$.  The scalar equations are immediately satisfied under this requirement, as both terms vanish identically. 

Remarkably, the scalar equation \eqref{eq:scalar.complex} is still satisfied {\em for any form of the Kähler potential}, not necessarily preserving the imaginary shift symmetry of our starting example \eqref{eq:shift.symmetry}. This result further generalizes the class of NLSMs included in our construction, showing that the appearance of bumpy black holes is a genuine feature of General Relativity coupled with any non-linear sigma model and 
a generic prediction of many supergravities \cite{Freedman:2012zz}. 

Notice that, when $\Psi$ is analytic, we can write the inhomogeneous Liouville equation \eqref{eq:liouville} in complex coordinates in a much simpler form
\begin{equation}
    2\partial_\zeta\partial_{\bar\zeta}P+\gamma e^{P(\zeta,\bar\zeta)}=-\kappa\,\partial_\zeta\partial_{\bar\zeta}K\ .
    \label{eq:liouville.complex}
\end{equation}
This allows for an simple solution in the particular case $\gamma=0$. Indeed, in that case the equation can be immediately integrated to
\begin{equation}
    P(\zeta,\bar\zeta)=-\frac12K(\Psi(\zeta),\bar\Psi(\bar\zeta))+\Xi_1(\zeta)+\bar\Xi_2(\bar \zeta)\ ,
\end{equation}
where $\Xi_{1,2}$ are arbitrary analytic functions reflecting the residual coordinate invariance on the conformal gauge. In the $\gamma\neq0$ case, the solution would depend on the details of the Kähler potential and on the specific analytic function $\Psi(\zeta)$.

\section{Including additional sources}
\label{sec:additional.sources}

In this section, we generalize the previous result including additional matter sources in Einstein equations, including additional scalar fields, Maxwell fields, and a perfect fluid. 

\subsection{Adding more scalar fields}
\label{sec:scalars}
A straightforward  generalization of the above presented construction is to add an arbitrary number of additional complex scalars $\Psi^A$ with no potential and a Kähler metric as in \eqref{eq:action.SuGra}. The important point is that the Kähler metric is diagonal in field space, it does not mix the different scalars $K_{\Psi^A\bar\Psi^B}=0$ for $A\neq B$. These fields contribute to the stress-energy tensor in the same way as the single field of the previous section. Thus, they decouple from the $r,t$ sector when we impose the analyticity condition in the complex variable $\zeta$, contributing only to the inhomogeneity of the Liouville equation for the function $P$. As we will show bellow in Section \ref{sec:higher.dimensions}, such fields are needed to generalize our bumpy solutions to higher dimensions. 

It is evident from the above construction, that we can add an arbitrary number of complex scalar fields $\phi^A$, as long as they depend on the $r,t$ coordinates only. These new scalars are allowed to be coupled through an interaction potential $V(\phi,\bar\phi)$ and an arbitrary NLSM metric $G_{AB}(\phi,\bar\phi)$. The corresponding stress-energy tensor reads
\begin{eqnarray} 
T_{\mu\nu}^{\sf Scalars}
=
G_{A\bar B}(\phi,\bar\phi)\,\partial_{\mu}\phi^{A}\,\partial_{\nu}\bar\phi^{\bar B}
-
g_{\mu\nu}\left(
\frac{1}{2}\,G_{A\bar B}(\phi,\bar\phi)\,g^{\rho\sigma}\partial_{\rho}\phi^{A}\partial_{\sigma}\bar\phi^{\bar B}
+
V(\phi,\bar\phi)
\right)\, . 
\end{eqnarray}
For radially dependent scalars with harmonic time dependence $\phi^A=e^{i\omega t} \varphi^A(r)$, this implies that the Einstein equations are modified as
\begin{align}
    &(tt)&&\cdots\ =\ \cdots\ -\ \kappa\, G_{A\bar B}\,\omega^2\varphi^A\bar\varphi^{\bar B}+\kappa\,fN^2\left( \frac12G^{AB}\left(\omega^2\varphi^A\bar\varphi^{\bar B}-\frac1f\varphi'^A\bar\varphi'^{\bar B}\right)+V\right)\ ,
    \nonumber\\    
    &(rr)&&\cdots\ =\ \cdots\ \ \kappa\, G_{A\bar B}\, \varphi'^A\bar\varphi'^{\bar B}-\frac1f\left(  \frac12G^{AB}\left(\omega^2\varphi^A\bar\varphi^{\bar B}-\frac1f\varphi'^A\bar\varphi'^{\bar B}\right)+V\right)\ ,
    \nonumber\\
    &(xx)&&
    \cdots\ =\ \cdots\ + \kappa\,r^2e^P\left(\frac12G^{AB}\left(\omega^2\varphi^A\bar\varphi^{\bar B}-\frac1f\varphi'^A\bar\varphi'^{\bar B}\right)+V\right)\ ,
    \nonumber\\
       &(xy)&&
    \cdots\ =\ \cdots\  \ ,
\end{align}
where the dots $\cdots$ stand for the terms which were already present in the equations in the absence of these new additional sources, as written in \eqref{eq:Einstein.vaccuum}. It is evident that the additional terms will only enter into the equations for $f$, and they do so in exactly the same form as in the vaccum case. Then any previously known hairy black hole, scalar cloud, or boson star solutions are recovered in this setup, but now with a bumpy transverse geometry.

\subsection{Including a Maxwell term}
 
The extension of the previous results to the presence of a Maxwell field is straightforward as, unlike what happens in the Pionic case, the Maxwell gauge field does not couple directly to the Sigma Model (this is the usual case in the cosets model appearing in SUGRA).  
The additional Maxwell equations read
\begin{equation}
\nabla_\mu F^{\mu\nu}=0\ ,
\end{equation}%
where we have assumed that the scalars on the coset space do not couple directly to the vector potential $A_\mu$. From the point of view of the section \ref{sec:Kahler}, this implies that we are focusing in {\em ungauged} supergravities. 
In our metric Ansatz  \eqref{eq:metric.Ansatz}, these equations are immediately solved by the static dyonic configuration 
\begin{equation}
F=\frac{q_e}{r^2}dt\wedge dr+q_me^{P}dx\wedge dy\ .
\label{eq:Maxwell.solution}
\end{equation}
It can be easily checked that $dF=0$ identically. This is natural, since we have a combination of the field produced by an electric source, and a magnetic part which is proportional to the volume form of $\Sigma_2$. Here $q_e$ and $q_m$ are integration constants with obvious interpretations as the electric and magnetic charges. 

On the other hand, Einstein equations \eqref{einsteinequations} have to be supplemented by the Maxwell energy-momentum tensor
\begin{equation}
T^{\sf Maxwell}_{\mu\nu}=F_{\mu\rho}F_{\nu}^{\ \rho}-\frac{1}{4}g_{\mu\nu}F_{\rho\tau}F^{\rho\tau}\ ,
\end{equation}
implying that the get the additional terms
\begin{align}
    &(tt)&&\ \cdots \ = \ \cdots \ +  \frac {\kappa f}{2r^4} \, \left( q_e^2+  N^2q_m^2\right)\ ,
        \nonumber\\
    &(rr)&&\ \cdots \ = \ \cdots \ -  \frac{\kappa}{2r^4fN^2}  \left( q_e^2+  N^2q_m^2\right)\ ,
        \nonumber\\
        &(xx)&&\ \cdots \ = \ \cdots \ +  \frac{\kappa e^P}{2r^2N^2} \left(  {q_m^2}+N^2{q_e^2}\right)\ ,
    \nonumber\\
        &(xy)&&\ \cdots \ = \ \cdots\ .   
\end{align}
Again, the new contributions to the Einstein equations do not spoil the separability of the $r,t$ and $x,y$ variables. 
This implies that, when the scalar fields satisfy the Cauchy-Riemann relations \eqref{eq:Cauchy-Riemann} they decouple from Einstein equations, leaving us with the standard (not bumpy) case, where the remaining equations for the metric are now coupled to a Maxwell field. The solution is then straightforwardly written, as
\begin{equation}
f(r)=\gamma-\frac{2m}{r}+\frac{q_e^2+q_m^2}{r^2}-\frac{\Lambda}{3}r^2\ ,
\qquad\qquad\qquad N(r)=1
\ .
\end{equation}
This is the blackening factor of the well-known asymptotically (A)dS or flat, charged magnetized topological black hole. But as in the uncharged case, the horizon manifold has a non-constant curvature in general, its geometry being determined by the matter fields via the inhomogeneous Liouville equation \eqref{eq:liouville.complex}.

It is interesting to notice that if the additional radially dependent scalars of the previous section $\Phi^A$ are charged, the right hand side of Einstein equations would not pick additional $x,y$ dependence {\em as long as the overall magnetic charge is zero.} This allows us to recover the charged hairy black hole solutions,  which in AdS represent superconducting holographic phases, but now with a horizon which is not  maximally symmetric.
\subsection{Coupling to a perfect fluid}
A similar construction to that of the previous sections can be performed by adding a perfect fluid to the matter content, whose stress-energy  tensor takes the form
\begin{eqnarray}
    T_{\mu\nu}^{\sf Fluid}=(p+\rho)u_\mu u_\nu+p\,g_{\mu\nu} \ .
\end{eqnarray}
In terms of a (unitary and time-like) four-velocity $u^\mu$, a pressure $p$, and an energy density $\rho$, the latest related by an equation of state which might contain the local temperature $p=p(\rho, T)$. This matter content is interesting as it describes both the normal fluid component and the superfluid component (corresponding to the BPS vortices) which, in many situations, can coexist. The corresponding fluid equations of motion are obtained from the conservation of the stress-energy tensor, as
\begin{eqnarray}
    \nabla^\mu T_{\mu\nu}^{\sf Fluid}=0\ .
\end{eqnarray}
Establishing the Ansatz $u_\mu=(u_0(r),0,0,,0)$ with $p=p(r)$ and $\rho=\rho(r)$, this conservation equation becomes
\begin{equation}
   p'+ (\rho+p)\left(\frac{f'}{2f}+\frac {N'}N\right)=0\ ,
\end{equation}
which is the standard conservation equation for a self-gravitating perfect fluid. Regarding the Einstein equations, they are modified as
\begin{align}
    &(tt)&&\ \cdots \ = \ \cdots \ -\ \kappa N^2 f\rho\ ,
        \nonumber\\
    &(rr)&&\ \cdots \ = \ \cdots \ -\ \kappa \,\frac { p}f\ ,
        \nonumber\\
        &(xx)&&\ \cdots \ = \ \cdots \ - \ \kappa e^P r^2 p\ ,
    \nonumber\\
        &(xy)&&\ \cdots \ = \ \cdots  \  .
\end{align}
As we can see, the separation procedure can still be performed, the resulting radial equations corresponding to the standard Tollman-Oppenheimer-Volkoff system. The separability persists if we charge the fluid, as long as there is no magnetic field. This allows us to construct stars \cite{Schaffner-Bielich:2020psc}, dark matter haloes \cite{Crespi:2024mgt}-\cite{Crespi:2025ygl}, and holographic metals \cite{Hartnoll:2010gu}-\cite{Acito:2026gbe}, with a non-homogeneous transverse geometry. 

\section{Higher dimensional black holes and black strings}
\label{sec:higher.dimensions}
It is well-known that black holes in higher dimensions posses different properties than their four-dimensional counterparts. For example, there exist asymptotically flat black holes with non-spherical horizons as the black ring solution and  its generalizations (see \cite{Emparan:2008eg, Horowitz:2012nnc}). Moreover, enough evidence of violations of cosmic censorship have been obtained during the last decades, for non-fine-tuned initial data \cite{Lehner:2011wc, Andrade:2020dgc, Figueras:2022zkg}. Given the structure of the solutions we have found in this work, it is natural to wonder whether they exist in General Relativity in higher dimensions. 

In this section, we show that there are at least four different possibilities leading to black holes and black strings/branes with bumpy horizons, by suitably extending the dimension of both the spacetime and the target space. In dimension $d=2+2n$, considering $n$ field doublets $\Psi_i$ with $i=1,\ldots,n$, we can construct generalizations of the Schwarzschild-Tangherlini-(A)dS spacetime. The structure of the field equations also allow to extend the horizon by including a $p$-dimensional Einstein piece, whose metric is multiplied by the square of the areal coordinate, namely by $r^2$. The latter is achieved with the same field content as before, i.e. with $n$ scalar doublets. The $p$-dimensional Einstein part of the horizon can be flatten by the inclusion of $p$ massless scalars, which  gives rise to a $p+2n$ dimensional target space. Finally, we show that the four-dimensional bumpy black holes can be extended to higher dimensions in the form of black strings or black branes. When the cosmological constant is non-vanishing, the four-dimensional geometry becomes warped by a function of the extended directions, while in the absence of $\Lambda$ in higher dimensions, the product between the four-dimensional part and the extended flat directions becomes a direct product, as for General Relativity in vacuum. The existence of these solutions is non-trivial, since it is known that four-dimensional black holes with matter fields cannot be trivially extended to higher dimensions by a simple direct product, due to the structure of the energy-momentum tensor. The BPS scalars in the NLSM we considered, by-pass such obstructions.

In what follows,  we show the details of these four scenarios.

\subsection{Bumpy horizons with dimension \texorpdfstring{$2n$}{2n}}
\label{sec:hd.n.Kahlers}
To generalize our construction to higher even dimensions $d=2+2n$, resulting in bumpy horizons with dimension $2n$, a simple method is to include $n$ further scalars as in the first proposal of section \ref{sec:scalars}. Namely, one can start with the action
\begin{equation}
\label{eq:action.SuGraHD}
I_{\sf EH-NLSM}=\int d^{2+2n}x\sqrt{-g}\left(\frac{R-2\Lambda}{2\kappa}  -\frac{1}{2}\sum_{i=1}^n K^i_{\Psi_i\bar\Psi_i}(\Psi_i, \bar\Psi_i) \,\nabla^\mu\Psi_i\nabla_{\mu}\bar\Psi_i   \right)  \ ,
\end{equation}
where $K^i_{\Psi_i\bar\Psi_i}(\Psi_i, \bar\Psi_i)$ is the mixed derivative of the arbitrary $i$-th Kähler potential $K^i(\Psi_i, \bar\Psi_i)$ with $i=1,...,n$.
Then the Ansatz for the spacetime metric can be generalized to its higher dimensional version  
\begin{equation}
\label{eq:bhsinHD}
    ds^2=-f(r)dt^2+\frac{dr^2}{f(r)}+r^2 \sum_{i=1}^nd\Sigma_i^2 \ 
\quad\qquad\mbox{where}\quad\qquad
    d\Sigma_i^2=e^{P_i(\zeta_i,\bar{\zeta_i})}d\zeta_i d\bar{\zeta_i}\ ,
\end{equation}
hence we are considering horizons that are direct products of $n$ two-dimensional Euclidean manifolds, in the conformal gauge for each of them, and with coordinates $(\zeta_i,\bar\zeta_i)$ with $i=1 \dots n$. Notice that we set $N=1$, as we are not planning to include additional matter sources in this section. 

The resulting Einstein equations can the be written, 
\small
\begin{align}
    &(tt)&& \!\!-\!f^2\left(\frac{(d-2)}{2rf}f'+\frac{(d-2)(d-3)}{2r^2}+\frac \Lambda f \right)+\frac{f}{2r^2}\sum_{i=1}^nR_{\Sigma_i} = \frac{\kappa f}{r^{2}}
\sum_{i=1}^{n}e^{-P_{i}}
 K_{\Psi_{i}\bar{\Psi}_{i}}^{i}\partial_{\zeta 
_{i}}\Psi_{i}\partial_{\bar{\zeta}_{i}}\bar{\Psi}_{i} 
\ ,
        \nonumber\\
    &(rr)&&\!\!\frac{(d-2)}{2rf}f'+\frac{(d-2)(d-3)}{2r^2}+\frac \Lambda f-\frac{1}{2fr^2}\sum_{i=1}^nR_{\Sigma_i} = -\frac{\kappa }{fr^{2}}
\sum_{i=1}^{n}e^{-P_{i}}
 K_{\Psi_{i}\bar{\Psi}_{i}}^{i}\partial_{\zeta 
_{i}}\Psi_{i}\partial_{\bar{\zeta}_{i}}\bar{\Psi}_{i} 
\ ,
\nonumber\\
        &(\zeta_i\bar\zeta_i)&&\!\!\! \frac{ e^{P_{i} 
}}{2}\!\left(  \!
\frac{r^2f^{\prime\prime}}{2}\!+\!{\left(  d\!-\!3\right) r f^{\prime}}%
\!+\!\frac{\left(  d \!-\!3\right)  \left(  d\!-\!4\right)  f}{2} -\frac{1}{2 }\sum_{j\neq i}R_{\Sigma_{i}}\!+\!\Lambda r^2         \! \right) \!=\! -\frac{\kappa}{2} \!\sum_{i,j\neq i} 
 {e^{P_i\!-\!P_{j}}} K_{\Psi_{i}\bar{\Psi}_{i}}^{i}\partial_{\zeta
_{i}}\Psi_{i}\partial_{\bar{\zeta}_{i}}\bar{\Psi}_{i}     ,
    \nonumber\\
        &(\zeta_i\zeta_i)&&  \!\!0 =K_{\Psi_i \bar{\Psi}_i}
\partial_{\zeta_i} \Psi_i \, \partial_{\zeta_i} \bar{\Psi}_i
\, .   
    \nonumber\\
        &(\bar\zeta_i\bar \zeta_i)&&\!\!0 = K_{\Psi_i \bar{\Psi}_i}
\partial_{\bar{\zeta_i}} \Psi_i \, \partial_{\bar{\zeta_i}} \bar{\Psi}_i \ .   
\label{eq:Einstein.higher.dimensions}
\end{align}
\normalsize
where we have used the direct product structure of the $(d-2)$-dimensional manifold $\Sigma$ and also the fact that two-dimensional spaces have identically vanishing Einstein tensors. The sum 
runs over the Ricci scalars of 
the $i$-th manifold in the direct product $\Sigma=\Sigma_1\times\Sigma_2\times\cdots\times\Sigma_n$, each of which in the corresponding conformal gauges \eqref{eq:bhsinHD}
\begin{equation}
    R_{\Sigma_i}=-4e^{-P_i(\zeta_i,\bar\zeta_i)} \partial_{\zeta_i}\partial_{\bar\zeta_i}P_i 
\end{equation}

The Einstein equation along the $\zeta_i\zeta_i$ and $\bar{\zeta_i}\bar{\zeta_i}$ directions are very restrictive, and they are solved for arbitrary Kähler potentials by setting $\Psi_j$ a holomorphic function of the corresponding complex coordinate $\zeta_j$, namely $\Psi_j=\Psi_j(\zeta_j)$ and $\bar\Psi_j=\bar\Psi_j(\bar\zeta_j)$.

Now, of course the matter equations are $n$ 
copies of \eqref{eq:scalar.complex}, which in terms of the complex fields $\Psi_j=\Psi_j(\zeta_j,\bar\zeta_j)$ which depend only on the complex coordinates $\zeta_j,\bar\zeta_j$, reduce to
\begin{equation}
K^j_{\Psi_j\bar\Psi_j} \,{\partial_{\zeta_j}\partial_{\bar\zeta_j}\Psi_j}  +K^j_{\Psi_j\Psi_j\bar\Psi_j} \partial_{\zeta_j}\Psi_j\partial_{\bar\zeta_j}\Psi_j  = 0\ ,
\label{eq:scalars.complex}
\end{equation}
and their complex conjugate. These are are identically solved by the holomorphic Ansatz $\Psi_j(\zeta_j,\bar\zeta_j)=\Psi_j(\zeta_j)$.

Finally, separation of variables and consistency of these equations imply that the blackening factor takes the form
\begin{equation}
    f(r)=\gamma-\frac{2m}{r^{d-3}}-\frac{2\Lambda}{(d-1)(d-2)}r^2
    \label{eq:blackeining.higher.dimensions}
\end{equation}
while the profile functions $P_i(\zeta_i,\bar\zeta_i)$ of the two-dimensional manifolds $\Sigma_i$ in the product $\Sigma$, are determined by the matter NLSM fields $\Psi_i(\zeta_i)$ via the inhomogeneous Liouville equations
\begin{equation}
2\partial_{\zeta_i}\partial_{\bar\zeta_i} P_{i}    +\left(  d-3\right)
\gamma e^{P_{i}}  = -\kappa \,\partial_{\zeta_i}\partial_{\bar\zeta_i}K
\label{eq:many.Liouville}
\end{equation}
Notice that the constant $\gamma$ is the same for all the equations with $i=1,\ldots ,n$.
\subsection{Bumpy horizons with dimension \texorpdfstring{$2n+p$}{2n+p} with a \texorpdfstring{$p$}{p}-dimensional Einstein factor}
\label{sec:Einstein}
Still with a $2n$-dimensional target space for the NLSM, the solutions of the previous section can be extended to the following family
\begin{equation}
    ds^2=-f(r)dt^2+\frac{dr^2}{f(r)}+r^2\left(d\mathcal{N}_p^2+\sum_{i=1}^nd\Sigma_i^2\right)\
\end{equation}
where $d\mathcal{N}_p$ is the line element of an Einstein manifold. The compatibility of the Einstein equations implies that
the blackening factor takes the form \eqref{eq:blackeining.higher.dimensions} and  the functions $P_i$ determining the conformal factor of the submanifolds $\Sigma_i$ satisfy equations \eqref{eq:many.Liouville}, as in the previous case. On the other hand, the Ricci tensor of the Einstein submanifold  is fixed to
\begin{equation}
    (R_p)^{a}_{ \ b}=(d-3)\gamma\delta^a_b\ .
\end{equation}
which completes the construction. 

\subsection{Bumpy horizons with dimension \texorpdfstring{$2n+p$}{2n+p} with a \texorpdfstring{$p$}{p}-dimensional toroidal factor}
\label{sec:torus}

A different extension to higher dimensions is constructed by adding flat directions to the target space of the NLSM. In other words, let us consider now the action
\small
\begin{equation}
\label{eq:action.axions}
I_{\sf}=\int d^{2+2n+p}x\sqrt{-g}\left(\frac{R-2\Lambda}{2\kappa}  -\frac{1}{2}\sum_{i=1}^n K^i_{\Psi_i\bar\Psi_i}(\Psi_i, \bar\Psi_i) \,\nabla^\mu\Psi_i\nabla_{\mu}\bar\Psi_i   -\frac{1}{2}\sum_{I=1}^p\partial_\mu\psi_I\partial^\mu\psi_{I}\right)  \ ,
\end{equation}
\normalsize
where $\psi^I$ are $p$ new real scalar fields. 
The corresponding equations of motion accommodate solutions of the form
\begin{equation}
    ds^2=-f(r)dt^2+\frac{dr^2}{f(r)}+r^2\left(\delta_{IJ}dx^Idx^J+\sum_{i=1}^nd\Sigma_i^2\right)\
\end{equation}
with $I,J=1,...,p$. The scalar equations of motion are solved by linear functions along each of the flat directions, namely
\begin{equation}
\psi_I=c_{(I)} x^I\ ,    
\end{equation}
where no sum in $I$ is intended. The Einstein equations read
\small
\begin{align}
    &(tt)&&   \dots \ = \ \dots \ + \frac {\kappa f}{2r^2} \sum_Ic_{(I)}^2
\ ,
    \nonumber\\
    &(rr)&& \dots \ = \ \dots\ - \frac \kappa{2fr^2}\sum_Ic_{(I)}^2
\ ,
    \nonumber\\
    &(\zeta_i\bar\zeta_i)&& \dots \ = \ \dots \ - \frac {\kappa e^{P_i}}4\sum_Ic_{(I)}^2
\, ,
\nonumber\\ 
&(JJ)&& 
\frac{r^2f^{\prime\prime}}{2}\!+\!{\left(  d-3\right) r f^{\prime}}%
\!+\!\frac{\left(  d-3\right)  \left(  d-4\right)  f}{2}+\Lambda r^2 -\frac{1}{2 }\sum_{j\neq i}R_{\Sigma_{i}}  =
\nonumber\\ 
&&&\qquad\qquad\qquad\qquad\qquad\qquad
     =\kappa c_{\left(  J\right)  }^{2}  - \frac\kappa2\sum_I c_{\left(  I\right)  }^{2}  
-\frac\kappa2
\sum_{i=1}^{n}e^{-P_{i}}
 K_{\Psi_{i}\bar{\Psi}_{i}}^{i}\partial_{\zeta 
_{i}}\Psi_{i}\partial_{\bar{\zeta}_{i}}\bar{\Psi}_{i} 
  \ ,  
\end{align}
\normalsize 
where, as in the previous sections, the dots stand for the terms that were already present in \eqref{eq:Einstein.higher.dimensions}. Notice that we omitted the $(\zeta\zeta)$ and $(\bar\zeta\bar\zeta)$ equations, as they remain unchanged. These equations can be separated as before, leading to the inhomogeneous Liouville equations \eqref{eq:many.Liouville} for the functions $P_i(\zeta_i,\bar \zeta_i)$, and the standard higher dimensional blackening factor \eqref{eq:blackeining.higher.dimensions}, as long as the coefficients $c_{(I)}$ are fixed to
\begin{equation}
    c_{(I)}^2=-\frac{(d-3)\gamma}\kappa \ , \qquad\qquad\qquad \forall \,I\in\{1,\ldots , p\}\ ,
\end{equation}
and  of course, as before, the scalars in the curved part of target space must be arbitrary holomorphic functions of the respective projective coordinates, namely $\Psi_i=\Psi_i(\zeta_i)$.  
Notice that if we require the new scalars $\psi_I$ to be non-ghosts we must have $\gamma\leq 0$. When $\gamma$ vanishes, the scalars vanish and we recover the results of the previous section.

\subsection{Black strings and black branes in higher dimensions}

As previously mentioned, the analytic  construction of homogeneous black strings and p-branes —starting from a four-dimensional black hole with matter fields and a cosmological constant— is not always straightforward. Here, we demonstrate that such solutions can indeed be constructed in the present setup. To do so, we consider the $d=4+p$ dimensional theory (with $p>1$) given by the action
\begin{equation}\label{EHNLSM-hdim}
I_{\sf EH-NLSM}=\int d^{4+p}x\sqrt{-g}\left(\frac{R-2\Lambda}{2\kappa}  -\frac{1}{2}\Omega^2(  \chi)\left(\left(  \nabla
\chi\right)  ^{2}+   \left(
\nabla\phi\right)  ^{2}   \right)\right)  \ ,
\end{equation}

and propose a higher dimensional extension of \eqref{lametrica} given by  
\begin{eqnarray}
ds^2=-f(r)N^2(r)dt^2+\frac{dr^2}{f(r)}+r^2 e^{P(x,y)}(dx^2+dy^2)+\sum_{i=1}^{p}dz_{i}dz^{i} \ . 
\end{eqnarray}
with $p$ extended directions parametrized by the $z_{i}$ coordinates. 

Analyzing as in the previous section the separability of Einstein's equations, we obtain a blackening factor $f(r)$ and a lapse function $N(r)$ given by \eqref{eq:vaccum.solution}. However, the remaining field equations are now more constraining, leading us to the condition that the cosmological constant must be zero in order to obtain a consistent equation for the $P(x,y)$, which then takes the form  in \eqref{eq:liouville}. 

These solutions correspond to homogeneous black strings or $p$-branes with a black hole in the transverse section. The horizon of this black hole has non-constant curvature, which is independent of the extended directions and maintains the same form as in the four-dimensional scenario. This result is valid for an arbitrary number of extra dimensions $p$, as the field equations have the same functional form as in the four-dimensional case. 

It is also possible to construct a five dimensional ($d=4+1$) black string with the same matter content as in the previous example and a non-vanishing cosmological constant, provided the four-dimensional section of the metric is scaled by a function of $z$. To achieve this, we consider the following metric ansatz
 \begin{eqnarray}
    ds^2=b^2(z)\left(-f(r)dt^2+\frac{dr^2}{f(r)}+r^2 e^{P(x,y)}(dx^2+dy^2)\right) +dz^{2} \ . 
\end{eqnarray}
In this configuration, the matter field equations do no contain the warp-factor, implying that the scalar fields can be set to be independent of the extended direction $z$. Thus, the matter equations remain the same as in the four-dimensional scenario. Regarding the Einstein equations, they are solved by
\begin{equation}
 f(r)=\gamma  -\frac{2m}{r}-\frac{\lambda}{3}r^2\, ,
 \qquad\qquad\qquad\qquad
    b(z)=\sqrt{\frac{2\lambda}{\Lambda}}\sin\left({\sqrt{\frac{\Lambda}{6}}(z-z_0)}\right)\ ,
\end{equation}
where $\lambda>0$ and $z_0$ are constants of integration. 
Notice that, as the metric is periodic in the $z$ direction, this coordinate can be compactified on a circle of length $\sqrt{6/\Lambda}$.

The case of a negative cosmological constant $\Lambda<0$ can be obtained by analytic continuation, replacing $\sin(\dots)$ by $i\sinh(\dots)$ and $\lambda$ by $-\lambda$. This results in an infinite range for the additional coordinate $z\in \,(-\infty,+\infty)$ and an asymptotically AdS section. A different solution for $\Lambda>0$ can be obtained if we first shift the integration constant $z_0\to z_0+\sqrt{6/\lambda}(\pi/2)$ (thus changing the $\sin(\dots)$ in the warp factor by a $\cos(\dots)$) and then perform the analytic continuation $\cos(\dots)\to \cosh(\dots)$. This has a range $z\in[0,\infty)$ and corresponds to an asymptotically dS section. 
%
When the integration constant vanishes  $\lambda=0$, the warp-factor degenerates to an exponential, as usual. 
 
{An interesting byproduct of our analysis is that it leads to the following suggestion related to inhomogeneous black strings. In higher dimensions $(d\geq 5)$, the factorized structure of the metric allows for a rich thermodynamic and geometric landscape. Since the solution allows for independent holomorphic maps $\Psi_{i}(\zeta_{i})$ for each complex co-direction of the transverse space, one can construct anisotropic black branes. An intriguing issue (on which we hope to come back in the near future) is that these BPS bumpy scalar configurations could provide a mechanism to sustain stable non-uniform black strings  (perhaps avoiding the Gregory-Laflamme instability). This question is presently under investigation.}

\section{Time dependent solutions}
\label{sec:time.dependent}
\subsection{Anisotropic bumpy cosmology}

One may wonder whether the construction we have developed can be extended to include time dependence. We can try a first attempt with a cosmological setup in $3+1$ dimensions, writing the metric as
\begin{equation}
    ds^2=-dt^2+a^2(t)\left(\frac{dr^2}{1-kr^2}+r^2e^{P(x,y)}(dx^2+dy^2)\right) \ .
\end{equation}
We complete the matter content with a perfect fluid with four-velocity $u^\mu=(u^0(t),0,0,0)$, density $\rho=\rho(t)$ and pressure $p=p(t)$.

Thus, we are describing a cosmological fluid which coexists with superfluid pionic vortices. This is an extremely interesting matter field content as it encodes both the normal fluid and the superfluid components allowing to detect the cosmological role of BPS vortices.

The conservation of the fluid energy then results in the familiar constraint 
\begin{equation}
    \dot \rho+3H(p+\rho)=0
\end{equation}
where $H=\dot a/a$ is the Hubble parameter. Using this definition and the result $\ddot a=a(\dot H+H^2)$ we can write the Einstein equations as
\small
\begin{align}
    &(tt)&& 3\left(H^2+\frac k{a^2}-\frac\Lambda3\right) +\frac{R_{\Sigma_2}-2}{2r^2a^2} =  \kappa\, \rho+\frac{\kappa}{2r^2a^2}e^{-P}\Omega^2(\chi) \left((\partial\chi)^2+(\partial\phi)^2\right)
        \nonumber\\
    &(rr)&&\!\! \!-\frac{a^2}{1-kr^2}\left(2 \dot H\!+\!3H^2 \!-\!\Lambda +\frac{R_{\Sigma_2}-2(1-kr^2)}{2a^2r^2}  \right) =  \frac{\kappa\, a^2}{1-kr^2}\! \left(p-\!\frac{e^{-P}\Omega^2(\chi) \left((\partial\chi)^2+(\partial\phi)^2\right)}{2a^2r^2}\right)
        \nonumber\\
        &(xx)&&\!\!\!-e^{P}r^2(k+a^2(2\dot H+3H^2-\Lambda )) = \kappa \,e^Pr^2a^2p+\frac\kappa 2{\Omega^2(\chi)}\left((\partial_x\chi)^2+(\partial_x\phi)^2-(\partial_y\chi)^2-(\partial_y\phi)^2\right)
    \nonumber\\&(xy)&&
    0=\kappa\,\Omega^2\left(\partial_x\chi\partial_y\chi+\partial_x\phi\partial_y\phi\right)\nonumber
\end{align}
\normalsize
As we see, again the $(xy)$ equation imposes the Cauchy-Riemann conditions \eqref{eq:Cauchy-Riemann}, which in turn decouple the $(xx)$ equation. This last equation can then be integrated to the standard Friedmann equation
\begin{eqnarray}
H^2+ \frac k{a^2}-\frac\Lambda3=\frac\kappa3\rho  
\end{eqnarray}
When replaced into the $(tt)$ and $(rr)$ equations, we recover the inhomogeneous Liouville equation for the particular case $\gamma=1$, 

\subsection{2+1-dimensional solutions}
The cosmological spacetimes that we showed in the previous section to be compatible with the BPS sector of the NLSMs, suggest a natural extension to $2+1$-dimensional spacetimes. Starting with the metric ansatz
\begin{equation}
    ds^2=-dt^2+a^2(t)\, e^{P(x,y)}(dx^2+dy^2) 
\end{equation} 
and restricting the matter sector to contain only the scalar fields, the action \eqref{eq:action.chi} extended to three dimensions, leading to field equations that are solved by the scalar fulfilling the Cauchy-Riemann equations \eqref{eq:Cauchy-Riemann}, while appropriately choosing the separation constant in the $tt$ component of Einstein equations leads to
\begin{align}
    H^2+\frac\gamma {a^2}=\Lambda
\end{align}
while the function $P(x,y)$ satisfied \eqref{eq:liouville} as in all the previous cases.  

When the scalars vanish, one recovers the cosmological foliation of flat, de Sitter or anti-de Sitter space, depending on whether $\Lambda$ vanishes, is positive, or negative, respectively \cite{dosmasunoflat,dosmasunocurved}. 

\subsection{Bumpy Birkhoff's-like theorem} 
The presence of the scalar fields leads to the propagation of an $s$-wave, giving dynamics to the spacetime even on spherical symmetry. Nevertheless, a Birkhoff's-like theorem can be proven for our black holes of section \ref{sec:NLSM} if we assume the matter fields still depend only on the angular directions\footnote{A similar analysis on a family of scalar tensor theories in 2+1-dimensions leading to regular black holes has been performed in \cite{Bueno:2025dqk}. There, the scalar is proportional to the circular coordinate $\phi$.}. For the metric we keep the ansatz \eqref{eq:metric.Ansatz}, but promoting the functions $f$ and $N$ to functions of $r$ and $t$. The Einstein equation along $(xy)$ direction imposes the Cauchy-Riemann relations \eqref{eq:Cauchy-Riemann}. With this, the equation in the $(tr)$ direction reads
\begin{align}
&(tr)&&    -\frac{\partial_tf}{rf}=0& \ \ \ \ \ \
\end{align}
implying that $\partial_t f=0$. The $(tt)$ and $(rr)$ components can then be written as in \eqref{eq:Einstein.vaccuum}, resulting in the solution \eqref{eq:vaccum.solution} for $f$ after separation of variables, together with the requirement $N=N(t)$. This last can be transformed into $N=1$ by a redefinition of the time variable. These equations also imply the inhomogeneous Liouville equation \eqref{eq:liouville} for the transverse conformal factor.

\section{Conclusions}
\label{sec:conclusions}
In this work, we have studied general two-dimensional non-linear sigma models (NLSMs) supporting static, gravitating solutions of General Relativity\footnote{See \cite{Anabalon:2012ta} for a general analysis of the compatibility of General Relativity with a gravitating NLSM in the ansatz that accommodates the Plebanski-Demianski solution, with scalars depending on the standard $(p,q)$ coordinates of the latter, namely on what would be the $(r,\theta)$ coordinates in our ansatz.}. Our solutions can be understood as generalizations of those constructed in \cite{Bardoux:2012aw} (see also \cite{Andrade:2013gsa}) for planar horizons and flat target space with scalars being linear, therefore trivially harmonic, on the two-dimensional horizon, as well as extensions of the recent bumpy black holes discovered in \cite{Canfora:2026col} in the context of gravitating Pions. The former, turned out to be relevant for the study of holographic conductivity in a setup that breaks translational invariance in the dual theory, and therefore leads to a finite DC-conductivity since the inhomogeneities induced by the scalar depending on the spatial boundary coordinates provide a mechanism for momentum relaxation \cite{Andrade:2013gsa}. The scalar fields that we have considered here can be used to construct lattices providing a natural, more general mechanism to introduce a breaking of the translational symmetry in the boundary. While the perturbations of the scalar field equations lead to the Jacobi equation for the NLSM, the separability of the back-reacting, perturbed charged system seems to be a non-trivial problem. We expect to report on this problem in the near future.

{A further comment is in order. The scalar field configurations (due to their BPS /holomorphic nature) affect the corresponding geometries making them ``bumpy''. One of the main points behind this choice can be seen from the AdS/CFT perspective: our formalism allows us to describe analytically inhomogeneous condensates in the boundary theory. This can deepen the present understanding of holographic conductivities.}
\section*{Acknowledgments}
We thank Andrés Anabalón for comments and Andrés Gomberoff and Aldo Vera for early work on this topic. F. C. has been funded by FONDECYT Grants
No. 1240048, 1240043 and 1240247 and
also supported by Grant ANID EXPLORACIÓN 13250014.  C. H appreciates the support of FONDECYT postdoctoral Grant 3240632. J.O. was supported by FONDECYT grant 1221504. N.G. is partially supported by CONICET grant PIP-2023-11220220100262CO, and
UNLP grant 2022-11/X931. N.  G. thanks Concepción U., San Sebastián U.,  Mayor U., and CECS for hospitality during the early stage of this work.

\end{document}